\documentclass[sigconf,screen,nonacm,display]{acmart}

\acmConference[ASE '26]{41st IEEE/ACM International Conference on Automated Software Engineering}{2026}{}
\acmYear{2026}
\acmDOI{}

\setcopyright{none}

\usepackage{booktabs}   
\usepackage{subcaption} 

\usepackage{paralist}
\usepackage[ruled,algosection,noend]{algorithm2e}
\usepackage{multirow}
\usepackage{bussproofs}
\usepackage{varwidth}
\usepackage{graphicx}
\usepackage{listings}
\usepackage{float}
\usepackage{multirow}
\usepackage[noend]{algorithmic}
\usepackage{mathpartir}
\usepackage{url}
\usepackage{hyperref}
\usepackage[capitalise]{cleveref}
\usepackage{xspace}
\usepackage{verbatim}
\usepackage{lipsum}
\usepackage{wrapfig}
\usepackage{ifsym}

\usepackage[inline, shortlabels]{enumitem}
\usepackage{url}
\usepackage{fancyvrb}
\usepackage[figurename=Fig.]{caption}
\usepackage{nanoc}
\usepackage{relsize}
\usepackage{marvosym}
\usepackage{microtype}
\usepackage{adjustbox}
\usepackage{multicol}

\usepackage{xcolor}
\usepackage{tikz}
\usepackage{tcolorbox}
\newtcolorbox{rqbox}{colback=gray!8, colframe=black!40, boxrule=0.4pt, arc=2pt, left=5pt, right=5pt, top=3pt, bottom=3pt, before skip=6pt, after skip=6pt}
\usepackage{array}
\usepackage{makecell}
\usepackage{pifont}

\usepackage[frozencache,cachedir=minted-cache]{minted}
\newminted[leancode]{lean4}{xleftmargin=5pt,fontsize=\footnotesize,escapeinside=||,linenos=true,numbersep=5pt}
\newminted[leancodesmall]{lean4}{xleftmargin=5pt,fontsize=\small,escapeinside=||,linenos=true,numbersep=5pt}
\newminted[leancodey]{lean4}{xleftmargin=5pt,fontsize=\footnotesize,escapeinside=||,linenos=true,numbersep=0pt}
\newminted[leancodex]{lean4}{xleftmargin=0pt,fontsize=\small,escapeinside=||,linenos=false,numbersep=5pt}
\newminted[leancodez]{lean4}{xleftmargin=0pt,fontsize=\footnotesize,escapeinside=||,linenos=false,numbersep=5pt}
\newminted[leansmall]{lean4}{xleftmargin=0pt,fontsize=\scriptsize,escapeinside=||,linenos=true,numbersep=0pt}
\usepackage{mathtools} 

\usetikzlibrary{arrows}
\usetikzlibrary{shapes.arrows}
\usetikzlibrary{shapes}
\usetikzlibrary{calc}
\usetikzlibrary{positioning}
\usetikzlibrary{fit}
\usetikzlibrary{backgrounds}
\usetikzlibrary{tikzmark}
\usetikzlibrary{snakes}

\usepackage{tikzsymbols} 

\tikzstyle{arrow}+=[thick,rounded corners=0.5em]
\tikzstyle{every picture}+=[remember picture,baseline]

\usepackage{multirow}
\usepackage{hhline}

\hypersetup{colorlinks,
  linkcolor=ACMDarkBlue,
  citecolor=ACMPurple,
  urlcolor=ACMDarkBlue,
  filecolor=ACMDarkBlue}

\floatname{algorithm}{Algorithm}

\SetAlFnt{\small}
\SetAlCapFnt{\small}
\SetAlCapNameFnt{\small}
\SetAlCapHSkip{0pt}
\IncMargin{-\parindent}

\floatstyle{plaintop}
\restylefloat{table}

\makeatletter 
\def\arcr{\@arraycr}
\makeatother

\definecolor{shadecolor}{gray}{1.00}
\definecolor{darkgray}{gray}{0.30}
\definecolor{violet}{rgb}{0.56, 0.0, 1.0}
\definecolor{forestgreen}{rgb}{0.13, 0.55, 0.13}
\definecolor{mygray}{rgb}{0.5,0.5,0.5}

\definecolor[named]{ACMBlue}{cmyk}{1,0.1,0,0.1}
\definecolor[named]{ACMYellow}{cmyk}{0,0.16,1,0}
\definecolor[named]{ACMOrange}{cmyk}{0,0.42,1,0.01}
\definecolor[named]{ACMRed}{cmyk}{0,0.90,0.86,0}
\definecolor[named]{ACMLightBlue}{cmyk}{0.49,0.01,0,0}
\definecolor[named]{ACMGreen}{cmyk}{0.20,0,1,0.19}
\definecolor[named]{ACMPurple}{cmyk}{0.55,1,0,0.15}
\definecolor[named]{ACMDarkBlue}{cmyk}{1,0.58,0,0.21}
\definecolor[named]{PaleGreen}{RGB}{196, 255, 231}
\definecolor[named]{PaleOrange}{RGB}{255, 213, 169}
\definecolor{intnull}{RGB}{213,229,255}

\definecolor{shadecolor}{gray}{1.00}
\definecolor{ddarkgray}{gray}{0.85}
\definecolor{darkgray}{gray}{0.30}
\definecolor{light-gray}{gray}{0.91}

\definecolor{ipurple}{HTML}{E5D8F3}
\definecolor{igreen}{HTML}{C2F4CF}
\definecolor{ired}{HTML}{AE0428}
\definecolor{iblue}{HTML}{3E74D1}
\definecolor{ilgray}{HTML}{E3E3E3}
\definecolor{idgray}{HTML}{747474}

\newcommand{\ie}{\emph{i.e.}\xspace}

\newcommand{\eg}{\emph{e.g.}\xspace}

\newcommand{\etal}{\emph{et~al.}\xspace}

\newcommand{\aka}{\textit{a.k.a.}\xspace}
\newcommand{\cf}{\textit{cf.}\xspace}

\newcommand{\wrt}{\emph{w.r.t.}\xspace}

\newcommand{\tname}[1]{\textsf{#1}\xspace}

\newcommand{\coq}{\tname{Rocq}}
\newcommand{\lean}{\tname{Lean}}
\newcommand{\K}{$\mathbb{K}$\xspace}

\newcommand{\ivy}{\tname{Ivy}}

\newcommand{\loom}{\tname{Loom}}
\newcommand{\dafny}{\tname{Dafny}}
\newcommand{\viper}{\tname{Viper}}
\newcommand{\verus}{\tname{Verus}}
\newcommand{\velvet}{\tname{Velvet}}
\newcommand{\tool}{\tname{LeetProof}}
\newcommand{\leetcode}{\tname{LeetCode}}
\newcommand{\verina}{\tname{VERINA}}
\newcommand{\clever}{\tname{CLEVER}}
\newcommand{\aristotle}{\tname{Aristotle}}
\newcommand{\plausible}{\tname{Plausible}}

\newcommand{\veil}{\tname{Veil}}
\newcommand{\fstar}{F${^{\star\!}}$\xspace}

\definecolor{pblue}{rgb}{0.13,0.13,1}
\definecolor{pgreen}{rgb}{0,0.5,0}
\definecolor{pred}{rgb}{0.9,0,0}
\definecolor{pgrey}{rgb}{0.46,0.45,0.48}

\definecolor{ckeyword}{HTML}{7F0055}
\definecolor{ccomment}{HTML}{3F7F5F}
\definecolor{cnumber}{HTML}{2A0099}

\newcommand{\code}[1]{\mintinline[fontsize=\small,escapeinside=||]{lean}{#1}}

\makeatletter
\protected\def\ccell#1#{%
  \kern-\fboxsep
  \@ccell{#1}%
}
\def\@ccell#1#2#3{%
  \colorbox#1{#2}{#3}%
  \kern-\fboxsep
}
\makeatother

\newcommand{\pts}{\mapsto}

\newcommand{\sep}{\ast}

\newcommand{\mcode}[1]{{\text{\code{#1}}}}

\definecolor{kwcolor}{HTML}{214A87}

\newcommand{\lw}[1]{\textcolor{kwcolor}{{\textbf{{#1}}}}}
\newcommand{\bdef}{\lw{bdef}}
\newcommand{\returns}{\lw{returns}}
\newcommand{\ensures}{\lw{ensures}}
\newcommand{\require}{\lw{require}}
\newcommand{\invariant}{\lw{invariant}}
\newcommand{\decreasing}{\lw{decreasing}}
\newcommand{\throw}{\lw{throw}}
\newcommand{\whilel}{\lw{while}}

\makeatother

\NewDocumentCommand{\hlt}{O{white} O{0pt} m m}{%
  \tikz[baseline=(X.base)]
    \node[
      fill=#3!30,
      inner sep=#2,
      anchor=base west,
      draw=#1,
      line width=0.1pt,
    ] (X) {#4};
}

\definecolor{diffred}{rgb}{1.0, 0.9, 0.9}   
\definecolor{diffgreen}{rgb}{0.9, 1.0, 0.9} 
\definecolor{highlightsnippet}{rgb}{1.0, 1.0, 0.85} 

\usepackage{color}
\definecolor{keywordcolor}{rgb}{0.7, 0.1, 0.1}   
\definecolor{tacticcolor}{rgb}{0.0, 0.1, 0.6}    
\definecolor{commentcolor}{rgb}{0.4, 0.4, 0.4}   
\definecolor{symbolcolor}{rgb}{0.0, 0.1, 0.6}    
\definecolor{sortcolor}{rgb}{0.1, 0.5, 0.1}      
\definecolor{attributecolor}{rgb}{0.7, 0.1, 0.1} 

\usetikzlibrary{matrix}
\usetikzlibrary{calc}

\setlist[itemize]{leftmargin=*}
\setlist[enumerate]{leftmargin=*}

\allowdisplaybreaks
\begin{document}

\newcommand{\mytitle}{Certified Program Synthesis with a Multi-Modal Verifier}

\newcommand{\runningtitle}{\mytitle}

\setlength\floatsep{1.25\baselineskip plus 3pt minus 2pt}
\setlength\textfloatsep{1.25\baselineskip plus 3pt minus 2pt}
\setlength\intextsep{1.25\baselineskip plus 3pt minus 2 pt}

\title[\runningtitle]{\mytitle}

\author{Yueyang Feng}
\authornote{Joint first authors.}
\affiliation{%
  \institution{National University of Singapore}
  \country{Singapore}
}
\email{yueyangfeng@u.nus.edu}
\orcid{0009-0002-7014-0159}

\author{Dipesh Kafle}
\authornotemark[1]
\affiliation{%
  \institution{National University of Singapore}
  \country{Singapore}
}
\email{dipesh@u.nus.edu}
\orcid{0009-0008-9368-4363}

\author{Vladimir Gladshtein}
\affiliation{%
  \institution{National University of Singapore}
  \country{Singapore}
}
\email{vovaglad@u.nus.edu}
\orcid{0000-0001-9233-3133}

\author{Vitaly Kurin}
\affiliation{%
  \institution{Neapolis University Pafos}
  \country{Cyprus}
}
\email{v.kurin@nup.ac.cy}
\orcid{0009-0007-9614-8892}

\author{George P\^{\i}rlea}
\affiliation{%
  \institution{National University of Singapore}
  \country{Singapore}
}
\email{gpirlea@u.nus.edu}
\orcid{0009-0008-5378-2815}

\author{Qiyuan Zhao}
\affiliation{%
  \institution{National University of Singapore}
  \country{Singapore}
}
\email{zhaoqiyuan@u.nus.edu}
\orcid{0000-0002-1017-1562}

\author{Peter M\"{u}ller}
\affiliation{%
  \institution{ETH Z\"{u}rich}
  \country{Switzerland}
}
\email{peter.mueller@inf.ethz.ch}
\orcid{0000-0001-7001-2566}

\author{Ilya Sergey}
\affiliation{%
  \institution{National University of Singapore}
  \country{Singapore}
}
\email{ilya@nus.edu.sg}
\orcid{0000-0003-4250-5392}

\begin{abstract}

%
\emph{Certified program synthesis} (\aka \emph{vericoding}) is the process of
automatically generating a program, its formal specification, and a
machine-checkable proof of program/specification alignment from a task
description given in a natural language.
%
%
Two key challenges make vericoding difficult. First, specifications synthesised
from natural language descriptions are often either too weak to be meaningful or
too strong to be implementable, yet existing approaches lack systematic means to
detect and correct such defects. Second, the landscape of program verifiers used
to validate the results is fragmented: each tool supports a particular reasoning
mode---\emph{auto-active} (\eg, \dafny, \verus) or \emph{interactive} (\eg,
\coq, \lean)---with its own trade-off between automation and expressivity. This
forces every synthesis methodology to be tailored to a single verification
paradigm, limiting the class of tasks it can handle effectively.

%
We propose to overcome both challenges by structuring the certified synthesis
workflow in stages around a \emph{multi-modal verifier}---a single tool that
combines dynamic validation, automated proofs, and interactive proof scripting
within one foundational framework.
%
%
We realise this idea in \tool, a new agentic pipeline built on \velvet, a
multi-modal program verifier embedded in the \lean theorem prover.
Multi-modality enables \tool to validate generated specifications via randomised
\emph{property-based testing} before any code is synthesised, decompose the
synthesis task into sub-problems guided by verification conditions, and delegate
residual proof obligations to frontier AI provers specialised for \lean.
%
%
We evaluate \tool on an extensive benchmark suite derived from prior work on
certified synthesis. Our specification validation uncovers defects in existing
reference benchmarks, and \tool's staged pipeline achieves a significantly
higher rate of fully certified solutions than a single-mode baseline at the same
fixed budget---consistently across two different frontier LLM backends. 


\end{abstract}




\maketitle


%


\section{Introduction}
\label{sec:intro}

\emph{Certified program synthesis}, or
\emph{vericoding}~\cite{VericodingBenchmark25}, is the task of producing, from a
description in a natural language, a program together with a formal
specification and a machine-checkable proof that the program meets it.
This is inherently difficult because it weaves together two problems
that are hard in isolation---(a)~translating an informal intent into a precise
formal specification and (b)~proving that a synthesised implementation satisfies
it---and each feeds back into the other: a faulty specification dooms even a
correct program, while an inadequate proof strategy leaves a correct
specification unverified.
Recent advances in LLM-based code generation have made vericoding feasible by
enabling models to produce formal specifications and proof
scripts~\cite{MisuLM024,SunSPB24,ChakrabortyEBFF25,YangLMYCGHLLLYZ25,Cobblestone24},
while \emph{program verifiers} serve as trustworthy oracles that check the
results.

The landscape of modern program verifiers broadly splits into two families.
\emph{Auto-active} verifiers, such as \dafny~\cite{Leino10},
\viper~\cite{Mueller-al:VMCAI16}, \verus~\cite{verus}, and
\fstar~\cite{ChakrabortyEBFF25}, ask the programmer to annotate code with
pre/postconditions, assertions, and loop invariants, which an SMT
solver~\cite{deMoura-Bjorner:TACAS08,BarbosaBBKLMMMN22} then checks
automatically.
\emph{Interactive} provers, such as \coq~\cite{rocq2025} and
\lean~\cite{MouraKADR15} instead require the user to construct proofs, step by
step, using \emph{proof scripts}, offering full expressivity at the cost of
greatly increased manual effort.

A growing body of work has tackled vericoding for each individual verifier and
paradigm. 
On the auto-active side, LLM-based approaches have been developed for
\dafny~\cite{MisuLM024,SunSPB24,LoughridgeSACSSMAMT24,BaksysZBDKH25,Laurel25,DafnyPro26},
\verus~\cite{YangLMYCGHLLLYZ25,AggarwalPW25,ChenLLGYLMYDCYLXZ25}, and
\fstar~\cite{ChakrabortyEBFF25}. On the interactive side, analogous efforts
target \lean~\cite{SongYA24} and \coq~\cite{Cobblestone24,PALM24,Rango24}.
Dedicated benchmarks have accompanied each line of work:
\tname{DafnyBench}~\cite{LoughridgeSACSSMAMT24} for \dafny, and
\verina~\cite{YeYHKYS25}, \clever~\cite{ThakurLTSZZDYC25}, and
\tname{VeriBench}~\cite{VeriBench25} for \lean.

This fragmentation of the verifier landscape, and of the vericoding efforts
built atop it, is not merely an inconvenience. Because each approach is built
around the idioms of a single verifier, the resulting workflows, prompting
strategies, and feedback loops are deeply entangled with tool-specific details,
making it difficult to distil reusable principles that transfer across different
verifiers or LLM backends.
Moreover, \emph{validating} formal program specifications in most of these
pipelines is typically limited to human
inspection~\cite{MaL0XB25,YeYHKYS25,MisuLM024,VericodingBenchmark25}, with no
systematic way to detect whether a generated specification is too weak
(admitting incorrect implementations) or too strong (ruling out every correct
one).

The specification-quality problem is not hypothetical. During this
work, we discovered that roughly 10\% of the reference specifications in two
state-of-the-art benchmarks, \verina~\cite{YeYHKYS25} and
\clever~\cite{ThakurLTSZZDYC25}, are defective---they either under-constrain or
incorrectly express the intended properties of the expected output (more on that
in \autoref{sec:eval-specs}). This reinforces the need for automated
specification quality checks as a component of any vericoding pipeline.

In this work, we propose to address both the tool-fragmentation and the
specification-quality problems by designing the vericoding workflow around a
\emph{multi-modal verifier}---a single tool that supports dynamic validation
(testing), automated verification (SMT), and interactive proof scripting
within one \emph{foundational} framework---one whose reasoning principles are
themselves mechanically verified, rather than trusted axioms.
Traditional \emph{single-mode} approaches commit to one paradigm: auto-active
tools such as \dafny~\cite{Leino10} offer fast SMT-backed feedback but cannot
express proofs outside the solver's reach, while interactive provers such as
\lean are more expressive but provide less automation.
A multi-modal verifier combines both, letting the pipeline choose the most
effective mode at each step: automation dispatches routine obligations,
interactive mode handles harder cases, and testing serves as a fast oracle.

We implement this idea on top of \velvet~\cite{VelvetPaper26}, a new verifier
for imperative programs embedded as a library in \lean. \velvet is built on
\loom~\cite{loom-paper}, a general framework for foundational multi-modal
verification in \lean. It integrates SMT-based
automation~\cite{qian2025leanautointerfacelean4,mohamed2025leansmtsmttacticdischarging}
with interactive \lean proofs and supports
\tname{QuickCheck}-style~\cite{ClaessenH00,plausible,GoldsteinCDPH24}
property-based testing (PBT) of programs, specifications, and loop invariants.
Since \velvet programs are ordinary \lean programs, the entire \lean
ecosystem---type checker, automation tactics, and the \tname{Mathlib}
library~\cite{mathlib20}---is available at every stage, and tricky verification
goals can be delegated to frontier AI provers~\cite{Aristotle25}.

Working with a foundational yet executable language (\ie, \lean) also lets us
address the specification-quality problem. Lahiri~\cite{Lahiri24} proposed
\emph{symbolic specification testing}: using specifications and an SMT-based verifier to 
prove that concrete tests do not fail, without executing code. While
effective for \dafny, purely symbolic checking becomes prohibitively expensive in
\lean, where SMT covers a smaller fraction of proof obligations. What turned out
to work surprisingly well is \emph{randomised specification testing}---using
property-based testing~\cite{GoldsteinCDPH24} to validate specifications against
test cases instead of formal proofs. PBT is fast, requires no proof engineering,
and catches the same class of defects at a fraction of the cost. A specification
that fails is rejected \emph{before} any code synthesis, catching
under-specification early and cheaply.

Combining multi-modal verification with randomised specification testing, we
present \tool: an AI-assisted agentic pipeline for end-to-end vericoding that
decomposes synthesis into independently validated \emph{stages}---specification
generation, program/invariant synthesis, and last-mile proof construction.
%
%
This structure maps onto a well-studied agentic
architecture~\cite{ZhangRFR24,YangJWLYNP24}, in which independent AI systems
work in tandem with deterministic symbolic tools, allowing the pipeline to be
composed modularly and optimised for costs by fine-tuning its specific
components. It was only by experimenting with different modes at each stage that
we identified PBT as the most cost-effective approach for specification
validation---illustrating the benefit of the staged, multi-modal design.

\vspace{-3pt}

\paragraph{Contributions}

This work makes the following contributions:

\vspace{-2pt}

\begin{itemize}

\item \tool: the first agentic pipeline for end-to-end vericoding built around
  a foundational multi-modal verifier, combining testing, automated SMT-based
  proofs, and AI-assisted interactive proof scripting in a unified, staged
  synthesis pipeline (\autoref{sec:overview}).

\item A new \emph{testing infrastructure} for \lean-based vericoding
  (\autoref{sec:pbt-filter}): type class-based output mutation for specification
  completeness checking, bounded enumeration for existential quantifiers in
  verification conditions, and a meta-programmed harness for randomised testing
  of synthesised programs and invariants.

\item An evaluation of \tool \emph{specification inference}
  (\autoref{sec:eval-specs}) demonstrating that our PBT-based specification
  generator achieves 97.4\% semantic accuracy on the \verina
  benchmark~\cite{YeYHKYS25}, while randomised specification testing uncovers
  defects in ${\sim}$10\% of two published benchmark suites: \verina and
  \clever~\cite{ThakurLTSZZDYC25}.

\item A new benchmark of 50 imperative-style \tname{LeetCode} problems with
  complexity annotations, and an evaluation of \tool \emph{synthesis pipeline}
  (\autoref{sec:evaluation}) showing that:
  \begin{inparaenum}[(a)]
  \item the multi-modal \tool pipeline produces significantly more fully
    certified solutions than a single-mode \lean baseline at the same fixed
    budget (\autoref{sec:rq2});
  \item all partially verified \velvet solutions are dischargeable with
    additional interactive effort, confirming the correctness of the synthesised
    artefacts (\autoref{sec:rq3}); and
  \item these gains are consistent across different frontier LLM
    backends (\autoref{sec:rq4}).
  \end{inparaenum}

\end{itemize}

\vspace{-7pt}

\begin{figure}[t]
\centering
\setlength{\abovecaptionskip}{5pt}
\setlength{\belowcaptionskip}{-10pt}
\begin{minipage}[t]{0.95\linewidth}
\begin{leancodesmall}
def countDivisors (n: |$\mathbb{N}$|) : |$\mathbb{N}$| :=
  ((List.range (n + 1)).filter
    (fun d => d > 0 |$\wedge$| n % d = 0)).length

def isPrime (n: |$\mathbb{N}$|) : Prop := n > 1 |$\wedge$| countDivisors n = 2

|\lw{method}| IsNonPrime (n: |$\mathbb{N}$|) |\lw{return}| (result: Bool)
  |\lw{ensures}| result = true |$\leftrightarrow$| |$\neg$|isPrime n
  do
    if n |$\leq$| 1 then |\lw{return}| true
    let |\lw{mut}| i : |$\mathbb{N}$| := 2
    let |\lw{mut}| ret : Bool := false
    |\lw{while}| i * i |$\leq$| n
    |\lw{invariant}| |$\neg$|ret |$\leftrightarrow$| |$\forall$|d, 2 |$\leq$| d |$\wedge$| d < i |$\to$| n % d |$\neq$| 0
    |\lw{invariant}| i |$\geq$| 2 |$\wedge$| (i - 1) * (i - 1) |$\leq$| n
    do
      if n % i = 0 then ret := true
      i := i + 1
    |\lw{return}| ret
\end{leancodesmall}
\end{minipage}
\caption{A \velvet method for checking non-primality.}
\label{fig:isnonprime}
\end{figure}

\begin{figure*}[t]
    \centering
    \begin{subfigure}[t]{\textwidth}
        \centering
        \begin{tcolorbox}[
          colback=purple!5,
          colframe=purple!40,
          arc=6pt,
          boxrule=0.5pt,
          left=6pt, right=6pt, top=4pt, bottom=4pt,
          width=0.95\linewidth
        ]
        \normalsize
        Given an array \code{nums}, return \code{true} if the array was
        originally sorted in non-decreasing order, then rotated some number of
        positions (including zero). Otherwise, return \code{false}.
        There may be duplicates in the original array. An array \code{A} rotated by \code{x} positions results
        in an array \code{B} of the same length such that
        \code{B[i] == A[(i+x) |\%| A.length]} for every valid index~\code{i}.
        \end{tcolorbox}
        \setlength{\abovecaptionskip}{-2pt}
        \setlength{\belowcaptionskip}{5pt}
        \caption{LeetCode problem 1752 statement in plain English}
        \label{fig:running_example}
    \end{subfigure}

    
    \begin{minipage}[t]{0.48\textwidth}
        \vspace{0pt} 
        \begin{subfigure}[t]{\textwidth}
            \centering
            \begin{leancodesmall}
-- A "drop" is a strict decrease from an element to
-- its cyclic successor.
def isDrop (nums : Array Int) (i : Nat) : Prop :=
  nums.size > 0 |$\land$| i < nums.size |$\land$|
  nums[(i + 1) % nums.size]! < nums[i]!

-- A sorted-and-rotated array has at most one cyclic drop.
def rotSortedProp (nums : Array Int) : Prop :=
  nums.size |$\leq$| 1 |$\lor$|
  (|$\forall$| i j : Nat, isDrop nums i |$\to$| isDrop nums j |$\to$| i = j)

-- No extra assumptions are needed for this problem.
def precondition (nums : Array Int) : Prop := True

-- The returned boolean should exactly decide "rotSortedProp"
def postcondition (nums : Array Int) (result : Bool) :=
  result = true |$\leftrightarrow$| rotSortedProp nums
\end{leancodesmall}
\caption{Lean specification: auxiliary functions and pre/postconditions}
\label{fig:spec-synth}
\end{subfigure}
        
\vspace{5pt}
        
        \addtocounter{subfigure}{1} 
        \begin{subfigure}[t]{\textwidth}
            \centering
            \begin{leancodesmall}
-- One of the residual goals to prove interactively
theorem goal_2
  (nums : Array Int)
  (i_2 : Nat)
  (invariant_inv_bounds : i_2 |$\le$| nums.size)
  -- other invariants are omitted for brevity
  (if_pos : {k |$\in$| Finset.range i_2 |$\mid$| nums[(k + 1) % nums.size]!
      < nums[k]!}.card |$\le$| 1)
  (done_1 : nums.size |$\le$| i_2)
  : postcondition nums true := by sorry

|\lw{prove\_correct}| CheckSortedAndRotated by
  loom_solve <;> ...
  exact (goal_2 ...)
            \end{leancodesmall}
            \caption{Proof script and an example extracted residual goal}
            \label{fig:proof-script}
        \end{subfigure}
    \end{minipage}
    \hfill
    \begin{minipage}[t]{0.48\textwidth}
        \vspace{0pt} 
        \addtocounter{subfigure}{-2} 
        \begin{subfigure}[t]{\textwidth}
            \centering
            \begin{leancodesmall}[highlightlines={12-23,31-33}, highlightcolor=highlightsnippet]
|\lw{method}| CheckSortedAndRotated (nums : Array Int)
|\lw{return}| (result : Bool)
|\lw{require}| precondition nums
|\lw{ensures}| postcondition nums result 
do
  let n := nums.size
  if n |$\leq$| 1 then |\lw{return}| true
  else
    let |\lw{mut}| drops : Nat := 0
    let |\lw{mut}| i : Nat := 0
    |\lw{while}| i < n
      -- The loop index stays within bounds. |\hfill \textcolor{blue}{\ding{172}}|
      |\lw{invariant}| "inv_bounds" (i |$\le$| n)
      -- The array size used in the loop is fixed.
      |\lw{invariant}| "inv_n_def" (n = nums.size)
      -- Modulo indexing is safe because n > 0.
      |\lw{invariant}| "inv_n_pos" (n > 0)
      -- drops counts the cyclic decreases
      |\lw{invariant}| "inv_drops_count"
        (drops = (Finset.filter
          (fun k : Nat => nums[(k + 1) % n]! < nums[k]!)
          (Finset.range i)).card)
      |\lw{decreasing}| n - i
    do
      let a := nums[i]!
      let b := nums[(i + 1) % n]!
      if b < a then drops := drops + 1
      i := i + 1
    |\lw{return}| (drops |$\leq$| 1)

#|\lw{guard}| (CheckSortedAndRotated #[4, 1, 2, 3]).extract |\hfill \textcolor{blue}{\ding{173}}|
velvet_plausible_test CheckSortedAndRotated |\hfill \textcolor{blue}{\ding{174}}|
\end{leancodesmall}
\setlength{\abovecaptionskip}{4pt}
\caption{Executable \velvet implementation and tests}   
\label{fig:velvet-synth}
\end{subfigure}
\addtocounter{subfigure}{1} 
\end{minipage}

\setlength{\abovecaptionskip}{5pt}
\setlength{\belowcaptionskip}{-10pt}
\caption{\tool pipeline: from the task description in a natural language to formally specified and verified \velvet code.}
\label{fig:combined_running_example}
\end{figure*}

\begin{figure*}[t!]
\setlength{\abovecaptionskip}{2pt}
\setlength{\belowcaptionskip}{-10pt}
  \centering
\input{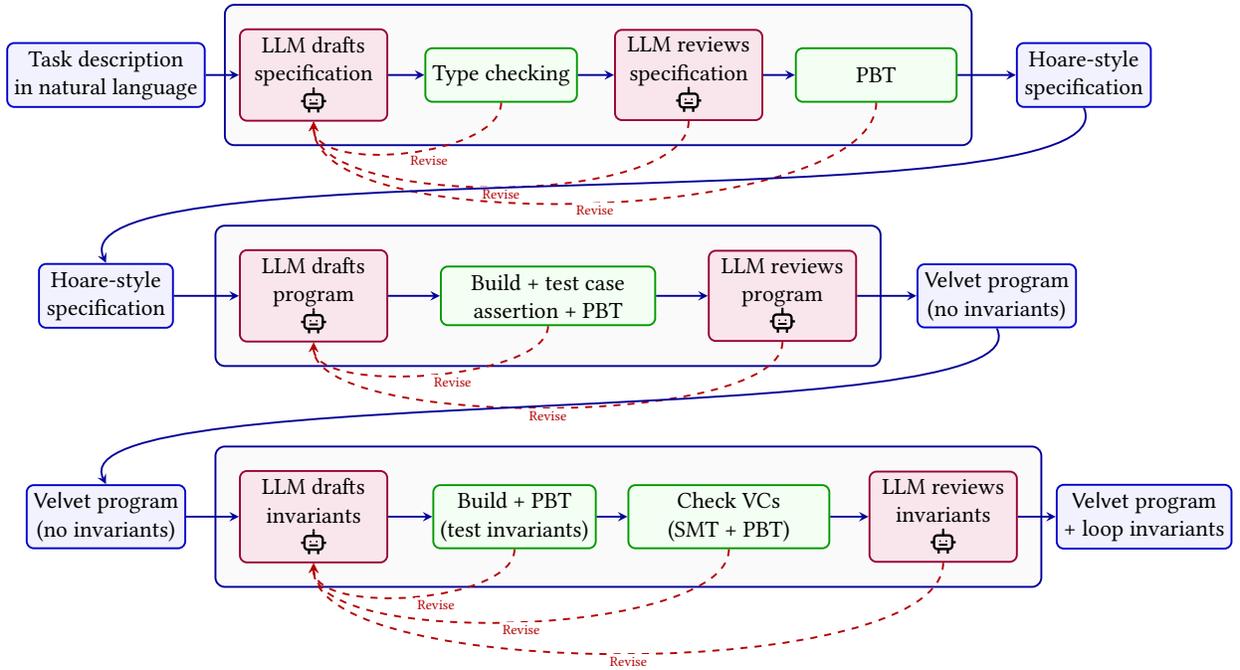}
\resizebox{0.92\textwidth}{!}{%
\begin{tikzpicture}[font={\fontsize{9.8}{12}\selectfont}, >=stealth]

  \def\rowy{0}

  \node[pipeline/artifact, minimum width=22mm] (in1) at (-0.3,\rowy)
    {Task description\\ in natural language};

  \node[pipeline/llmstep, minimum width=22mm] (draft1) at (2.8,\rowy)
    {LLM drafts\\ specification\\[-1pt]\llmboticon};
  \node[pipeline/step, minimum width=18mm] (typecheck) at (5.6,\rowy)
    {Type checking};
  \node[pipeline/llmstep, minimum width=22mm] (review1) at (8.4,\rowy)
    {LLM reviews\\ specification\\[-1pt]\llmboticon};
  \node[pipeline/step, minimum width=24mm] (tests1) at (11.2,\rowy)
    {PBT};
  \node[pipeline/artifact, minimum width=20mm] (out1) at (14.3,\rowy)
    {Hoare-style\\ specification};

  \begin{scope}[on background layer]
    \node[pipeline/panel, fit=(draft1) (review1) (tests1), inner xsep=6pt, inner ysep=10pt] {};
  \end{scope}

  \draw[pipeline/flow] (in1) -- (draft1);
  \draw[pipeline/flow] (draft1) -- (typecheck);
  \draw[pipeline/flow] (typecheck) -- (review1);
  \draw[pipeline/flow] (review1) -- (tests1);
  \draw[pipeline/flow] (tests1) -- (out1);

  \draw[pipeline/revise] (typecheck.south) .. controls +(0,-0.8) and +(0,-0.8) ..
    node[pipeline/label, pos=0.5, text=red!70!black, fill=white, inner sep=1pt] {\scriptsize Revise} (draft1.south);
  \draw[pipeline/revise] (review1.south) .. controls +(0,-1.3) and +(0,-1.3) ..
    node[pipeline/label, pos=0.5, text=red!70!black, fill=white, inner sep=1pt] {\scriptsize Revise} (draft1.south);
  \draw[pipeline/revise] (tests1.south) .. controls +(0,-1.8) and +(0,-1.8) ..
    node[pipeline/label, pos=0.5, text=red!70!black, fill=white, inner sep=1pt] {\scriptsize Revise} (draft1.south);

  \def\rowyy{-3.3}

  \node[pipeline/artifact, minimum width=20mm] (in2) at (-0.3,\rowyy)
    {Hoare-style\\ specification};

  \node[pipeline/llmstep, minimum width=22mm] (draft2) at (2.8,\rowyy)
    {LLM drafts\\ program\\[-1pt]\llmboticon};
  \node[pipeline/step, minimum width=32mm] (checks2) at (6.3,\rowyy)
    {Build + test case\\ assertion + PBT};
  \node[pipeline/llmstep, minimum width=22mm] (review2) at (9.8,\rowyy)
    {LLM reviews\\ program\\[-1pt]\llmboticon};
  \node[pipeline/artifact, minimum width=22mm] (out2) at (13.0,\rowyy)
    {\velvet program\\ (no invariants)};

  \begin{scope}[on background layer]
    \node[pipeline/panel, fit=(draft2) (review2), inner sep=10pt] {};
  \end{scope}

  \draw[pipeline/flow] (in2) -- (draft2);
  \draw[pipeline/flow] (draft2) -- (checks2);
  \draw[pipeline/flow] (checks2) -- (review2);
  \draw[pipeline/flow] (review2) -- (out2);

  \draw[pipeline/revise] (checks2.south) .. controls +(0,-0.8) and +(0,-0.8) ..
    node[pipeline/label, pos=0.5, text=red!70!black, fill=white, inner sep=1pt] {\scriptsize Revise} (draft2.south);
  \draw[pipeline/revise] (review2.south) .. controls +(0,-1.3) and +(0,-1.3) ..
    node[pipeline/label, pos=0.5, text=red!70!black, fill=white, inner sep=1pt] {\scriptsize Revise} (draft2.south);

  \draw[draw=blue!60!black, thick, ->]
    (out1.south) .. controls +(0.8,-1.8) and +(-0.8,1.8) .. (in2.north);

  \def\rowyyy{-6.6}

  \node[pipeline/artifact, minimum width=22mm] (in3) at (-0.3,\rowyyy)
    {\velvet program\\ (no invariants)};

  \node[pipeline/llmstep, minimum width=22mm] (draft3) at (2.8,\rowyyy)
    {LLM drafts\\ invariants\\[-1pt]\llmboticon};
  \node[pipeline/step, minimum width=24mm] (checks3) at (5.8,\rowyyy)
    {Build + PBT\\ (test invariants)};
  \node[pipeline/step, minimum width=30mm] (vcchecks) at (9.0,\rowyyy)
    {Check VCs\\ (SMT + PBT)};
  \node[pipeline/llmstep, minimum width=22mm] (review3) at (12.2,\rowyyy)
    {LLM reviews\\ invariants\\[-1pt]\llmboticon};
  \node[pipeline/artifact, minimum width=24mm] (out3) at (15.2,\rowyyy)
    {\velvet program\\ + loop invariants};

  \begin{scope}[on background layer]
    \node[pipeline/panel, fit=(draft3) (review3), inner sep=10pt] {};
  \end{scope}

  \draw[pipeline/flow] (in3) -- (draft3);
  \draw[pipeline/flow] (draft3) -- (checks3);
  \draw[pipeline/flow] (checks3) -- (vcchecks);
  \draw[pipeline/flow] (vcchecks) -- (review3);
  \draw[pipeline/flow] (review3) -- (out3);

  \draw[pipeline/revise] (checks3.south) .. controls +(0,-0.8) and +(0,-0.8) ..
    node[pipeline/label, pos=0.5, text=red!70!black, fill=white, inner sep=1pt] {\scriptsize Revise} (draft3.south);
  \draw[pipeline/revise] (vcchecks.south) .. controls +(0,-1.3) and +(0,-1.3) ..
    node[pipeline/label, pos=0.5, text=red!70!black, fill=white, inner sep=1pt] {\scriptsize Revise} (draft3.south);
  \draw[pipeline/revise] (review3.south) .. controls +(0,-1.8) and +(0,-1.8) ..
    node[pipeline/label, pos=0.5, text=red!70!black, fill=white, inner sep=1pt] {\scriptsize Revise} (draft3.south);

  \draw[draw=blue!60!black, thick, ->]
    (out2.south) .. controls +(0.8,-1.6) and +(-0.8,1.6) .. (in3.north);

\end{tikzpicture}
}
\caption{Partial \tool pipeline: specification generation (top), program synthesis
  (middle), and invariant inference (bottom). Solid blue arrows show artefacts
  flowing between stages; dashed red arrows show revision loops within each
  stage.}
\label{fig:pipeline-spec-and-program}
\end{figure*}

\section{Background}
\label{sec:background}

We start by briefly introducing the two systems that \tool builds on: the \lean
theorem prover and the \velvet verifier.

\vspace{-5pt}

\subsection{Lean}

\lean~\cite{MouraU21} is an open-source theorem prover and dependently typed
programming language. Its expressive type system allows users to state and prove
theorems (including statements about pure \emph{functional} programs)
interactively using \emph{proof scripts}. \lean's mathematical library
\tname{Mathlib}~\cite{mathlib20} contains over 210,000 formalised theorems,
making it one of the most extensive such libraries in any proof assistant. 
This rich ecosystem has made \lean the platform of choice for major AI-assisted
mathematical reasoning efforts, including
\tname{AlphaProof}~\cite{AlphaProof25}, which achieved silver-medal performance
at the 2024 International Mathematical Olympiad (IMO), and
\aristotle~\cite{Aristotle25}, which reached gold-medal level at the 2025 IMO.
Beyond mathematics, \lean also serves as a \emph{meta-verifier}: its hygienic
macro system and metaprogramming facilities~\cite{metalean} allow users to embed
domain-specific reasoning frameworks as libraries. 


\vspace{-5pt}

\subsection{Velvet}

\velvet~\cite{VelvetPaper26} is a Hoare-style~\cite{Hoare69} program verifier
for \emph{imperative} programs, embedded as a library in \lean via \loom
framework~\cite{loom-paper}. Programs in \velvet are annotated with
pre/postconditions and loop invariants, and \loom generates verification
conditions (VCs) whose validity implies program correctness. Because \loom is
itself formalised in \lean, this implication is a machine-checked
theorem---making \velvet a \emph{foundational} verifier that needs not be
trusted.

\autoref{fig:isnonprime} shows a complete \velvet example: a method that decides
whether a natural number is \emph{not} prime.
The specification relies on two auxiliary \lean definitions, \code{countDivisors}
and \code{isPrime} (lines~1--5). These are ordinary \lean functions that use
list filtering and quantification; they are executable but would be inefficient
to run on large inputs. This is a deliberate design choice: specifications
in \velvet are \lean propositions and \emph{need not} be executable.

The imperative method \code{IsNonPrime} (lines~7--19) comes with the
postcondition (\code{|\lw{ensures}|} clause at line~8), which states that its
Boolean result corresponds exactly to \code{|$\neg$|isPrime n}. The loop
(lines~13--18) carries two invariants: \code{ret} is false iff no divisor of
$n$ has been found in $[2, i)$, and $i \geq 2$ with $(i{-}1)^2 \leq n$.
\velvet programs are ordinary \lean programs (with monadically  encoded
effects~\cite{fstarmonads}): one can execute, \eg, \#\code{|\lw{eval}| (IsNonPrime
42).run} to test the method.

The \velvet command \code{|\lw{prove\_correct}|} triggers VC generation. For
this example, it produces 15~verification conditions---plain \lean theorems that
together imply correctness of \code{IsNonPrime} \wrt the ascribed specification.
Of these, 14 are discharged fully automatically with the help of SMT-based
automation (via \tname{lean-auto}~\cite{qian2025leanautointerfacelean4}) and
\lean tactics such as \code{grind}~\cite{grind} and \code{aesop}~\cite{aesop23}.
The single remaining VC requires an \emph{interactive} proof: it is the
number-theoretic fact stating that a number is prime if and only if it has no
divisors between~2 and its integer square root. This proof can be written
manually in \lean or delegated to an AI prover such as \aristotle.



\vspace{-5pt}

\section{A Tour of \tool}
\label{sec:overview}

\tool takes a programming task in natural language and produces a verified
\velvet program with a machine-checkable correctness proof in \lean. The task is
decomposed into three stages: specification (\autoref{sec:specsyn}), program
synthesis (\autoref{sec:progsyn}), and proof (\autoref{sec:proofsyn}); each
producing a validated intermediate artefact.
\autoref{fig:combined_running_example} illustrates the pipeline on \tname{LeetCode}
problem
1752:\footnote{\url{https://leetcode.com/problems/check-if-array-is-sorted-and-rotated/}}
the natural language statement (\autoref{fig:running_example}) becomes a formal
specification (\autoref{fig:spec-synth}), then a \velvet implementation
(\autoref{fig:velvet-synth}), and finally a correctness proof
(\autoref{fig:proof-script}).

\subsection{Specification Synthesis}
\label{sec:specsyn}

The first stage translates a natural-language problem description into a formal
\lean specification (\autoref{fig:spec-synth}). The LLM also generates several
concrete test cases alongside the specification; these serve both as validation
inputs and as human-readable documentation for the formal definitions. Since
every subsequent stage---program synthesis, invariant inference, and correctness
proof---depends on this specification, ensuring its quality is crucial.
An LLM-generated specification can go wrong in several ways: it may fail to
type-check, admit trivially correct implementations by being too weak, impose
constraints that no implementation can satisfy by being too strong, or simply
misinterpret the problem. Our pipeline addresses these risks in steps, as shown
in the top row of \autoref{fig:pipeline-spec-and-program}.

The process begins with an LLM proposing a draft specification together with
${\sim}$10 test cases (some derived from the problem statement, some are synthesised).
The draft must type-check in \lean; an LLM judge then reviews it for common
issues. If either check fails, the LLM revises the draft. To gain further
confidence, we apply our new take on \emph{randomised specification testing}
using \lean's \plausible PBT library~\cite{plausible}. For each test
case, \tool programmatically generates three checks: (a)~the test input
satisfies the \code{precondition}, filtering irrelevant inputs; (b)~the intended
input/output pair satisfies the \code{postcondition}, catching specs that reject
correct answers; and (c)~no alternative output satisfies the
\code{postcondition} for the same input, ruling out
under-specification~\cite{Lahiri24,BaksysZBDKH25}.

\autoref{fig:spec-pbt} illustrates this on our running example. The
specification defines a \code{precondition} (\ding{172}) requiring the array
size to exceed~1, and a \code{postcondition} (\ding{173}) stating that
\code{result = true} \emph{implies} the array is sorted-and-rotated. A test case
(\ding{174}) is generated alongside. Each PBT check is a \lean definition typed
as \code{Prop}---syntactically identical to a theorem. Normally this requires a
deductive proof; however, the tactic \code{plausible'} (a non-failing variant of
\lean's \plausible tactic~\cite{plausible}) instead searches for
counterexamples by generating random inputs. If a counterexample is found, the
check fails and the spec or test case is flagged. If none is found,
\code{plausible'} silently admits the goal (\ie, inserts a \code{sorry}),
letting the pipeline proceed. This is safe in terms of verification because
these checks are validation guards, not part of the final correctness
certificate.

In this example, checks \ding{175} and \ding{176} both pass: the input
\#\code{[1, 2, 3]} satisfies the precondition (size $3 > 1$), and the expected
output \code{true} satisfies the postcondition. However, the uniqueness
check~\ding{177} fails: PBT finds \code{result = false} as a counterexample.
Because the postcondition uses an implication ($\to$ rather than
$\leftrightarrow$), setting \code{result} to \code{false} makes it
\emph{vacuously true}---so the specification admits a degenerate implementation
that always returns \code{false}. This is exactly the class of
under-specification that randomised testing is designed to catch.
Beyond filtering, the generated tests serve as human-readable documentation for
the formal specification: a user can inspect the accepted concrete input/output
pairs to understand what the spec means without reading the formal \lean
definitions, closing an important gap between the informal task description and
the rigorous specification used in subsequent stages.
An alternative is \emph{symbolic specification testing}~\cite{Lahiri24}, which
\emph{proves} that each test passes under the specification via an SMT-based
verifier. We experimented with this and found it ineffective in \lean: even
simple proof obligations were expensive to construct, making PBT the clear
winner in cost.
%

\vspace{-5pt}

\subsection{Program and Invariant Synthesis}
\label{sec:progsyn}


\velvet is an \emph{intrinsic} program verifier: correctness proofs rely on
program-level annotations---most importantly, \emph{loop invariants}---that
drive the verification condition (VC) generation covered in
\autoref{sec:proofsyn}. Consequently, this stage couples two sub-tasks:
synthesising the program code and inferring its loop invariants.

\begin{figure}[t]
\centering
\begin{minipage}[t]{0.95\linewidth}
\begin{leancodesmall}
def precondition (nums : Array Int) : Prop := |\hfill \textcolor{blue}{\ding{172}}|
  nums.size > 1 -- not the final one, too restrictive
|\hfill \textcolor{blue}{\ding{173}}|
def postcondition (nums : Array Int) (result : Bool) :=
  result = true |$\to$| rotSortedProp nums
|\hfill \textcolor{blue}{\ding{174}}|
def test1_nums : Array Int := #[1, 2, 3]
def test1_Expected : Bool := true

def precondition_test1 : precondition test1_nums := |\hfill \textcolor{blue}{\ding{175}}|
  by simp; plausible'

def postcondition_test1 : |\hfill \textcolor{blue}{\ding{176}}|
  postcondition test1_nums test1_Expected := by
  simp; plausible'

-- counterexample to uniqueness: result = false |\hfill \textcolor{blue}{\ding{177}}|
def uniqueness_test1 (result : Bool) : 
   result |$\ne$| test1_Expected |$\to$|
  ¬ postcondition test1_nums result := by
  simp ; plausible'
\end{leancodesmall}
\end{minipage}
\setlength{\belowcaptionskip}{-10pt}
\caption{PBT checks for generated specifications and test cases.}
\label{fig:spec-pbt}
\end{figure}

\vspace{3pt}

\emph{Code generation.}~~
Given a validated specification from Stage~1, the LLM generates a candidate
\velvet program (middle row of \autoref{fig:pipeline-spec-and-program}). For our running
example, the result is shown in \autoref{fig:velvet-synth}. Because \velvet
programs are ordinary \lean programs, they can be tested immediately: \ding{173}
shows a concrete test case assertion (via \lean's \#\code{|\lw{guard}|} command), 
and \ding{174} invokes PBT via a helper macro. PBT checks the program against
both the specification and concrete test cases by running it on random inputs
satisfying the precondition and verifying the postcondition holds. 
Any failure is fed back to the LLM for revision. An LLM judge additionally
reviews the output for common issues (\eg, producing pure functional \lean code
rather than imperative \velvet code, which would bypass invariant inference and
defeat the purpose of multi-modal verification).


\paragraph{Invariant inference.}~~
Loop invariant inference---finding inductive properties that hold at every
iteration and are strong enough to imply the postcondition---is a notoriously
difficult problem, closely intertwined with the specification inference and
program correctness proof~\cite{FlanaganL01,Laurel25,BaksysZBDKH25}.
In auto-active verifiers like \dafny, candidate invariants can be checked
cheaply via SMT. In \lean, SMT covers a smaller fraction of obligations (though
it remains effective for linear arithmetic), so we again exploit multi-modality.

The invariant inference loop (bottom row of \autoref{fig:pipeline-spec-and-program})
proceeds as follows. The LLM proposes candidate invariants (\ding{172} in
\autoref{fig:velvet-synth}). \tool then triggers VC generation via \loom and
attempts to discharge the resulting VCs with automated tactics. For VCs that
remain, PBT searches for counterexamples. If a counterexample is found---or if
the LLM, inspecting the remaining VCs, judges one to be unprovable given the
current invariants---the feedback is propagated back and the LLM revises its
invariants.

\autoref{fig:program-pbt} illustrates this: replacing \code{=} with \code{<} in
the \code{inv_drops_count} invariant causes PBT to immediately produce a
counterexample:
\begin{minted}[breaklines=true, fontsize=\small]{text}
[velvet_plausible_test] FAIL: invariant "inv_drops_count" doesn't hold
nums = #[-18, 10, -13, 11, 8, 17, 13, -19, 15, -1, -27, -25]
\end{minted}
\noindent
The buggy invariant claims \code{drops} is \emph{strictly less} than the
number of cyclic decreases seen so far. For this input, the two quantities are
equal at some iteration, violating the strict inequality.

%
The \code{|\lw{decreasing}|} clause (line~23 of \autoref{fig:velvet-synth})
specifies a \emph{termination measure} for the loop, which is also inferred by
LLM at this stage and formally verified in the subsequent proof synthesis stage.
This clause is optional: \velvet supports both \emph{partial correctness} (the
postcondition holds \emph{if} the program terminates) and \emph{total
correctness} (the program terminates and satisfies its
postcondition)~\cite{loom-paper}. The choice can be configured per task in the
vericoding pipeline.

\vspace{-3pt}

\paragraph{Guarantees at the end of this stage.}

PBT can refute incorrect invariants but cannot prove that the surviving ones are
inductive or sufficiently strong. The goal of this stage is therefore
\emph{high-confidence} invariants: candidates that pass all automated checks
(PBT, SMT, LLM review) without yet having been formally proved correct. This
leaves room for incompleteness---an invariant may turn out to be too weak during
the proof stage---but in practice, the combination of PBT filtering and
LLM-based VC assessment produces invariants that rarely need revision
(\cf~\autoref{sec:evaluation}). A program that passes all checks is forwarded to
the next stage for the final formal proof.

\begin{figure}[t]
\centering
\begin{minipage}[t]{0.95\linewidth}
\begin{leancodesmall}
|\lw{invariant}| "inv_drops_count"
    (drops |\colorbox{diffred}{=}||\colorbox{diffgreen}{<}| (Finset.filter
       (fun k : Nat => nums[(k + 1) |\%| n]! < nums[k]!)
         (Finset.range i)).card)
\end{leancodesmall}
\end{minipage}
\setlength{\belowcaptionskip}{-10pt}
\caption{Incorrect invariant caught by PBT.}
\label{fig:program-pbt}
\end{figure}

\vspace{-6pt}

\subsection{Proof Synthesis and Residual Obligations}
\label{sec:proofsyn}

Once a \velvet program with high-confidence invariants is ready, the
\code{|\lw{prove\_correct}|} command triggers VC generation via \loom. Each VC is an
ordinary \lean theorem, which means we can dispatch it using \emph{any} method
available in the \lean ecosystem---a key advantage of working inside a
foundational proof assistant.

%
\tool first attempts to close every VC automatically using a combination of
SMT-based
tactics~\cite{qian2025leanautointerfacelean4,mohamed2025leansmtsmttacticdischarging}
and built-in \lean tactics such as \code{grind}~\cite{grind} and
\code{aesop}~\cite{aesop23}. For our running example, this step discharges 14
out of 18~VCs, leaving 4~\emph{residual obligations}.

%
A residual obligation is a VC that automation cannot close. In the listing,
these appear as \lean theorems with \code{sorry} placeholders
(\autoref{fig:proof-script}), which are then plugged into the
\code{|\lw{prove\_correct}|} block. The goal \code{goal_2} in
\autoref{fig:proof-script} is one such obligation: it requires showing that the
loop's postcondition follows from the invariants at termination---a general
mathematical fact that SMT cannot discharge.

To close residual obligations, \tool employs an LLM-based agent inspired by the
 decomposition technique of \tname{Hilbert}~\cite{varambally2025hilbert}. The
 prover agent has access to \tname{Mathlib}~\cite{mathlib20} search, can
 register auxiliary lemmas, and can decompose a goal into smaller sub-goals that
 it attempts recursively. 
 For particularly hard obligations, \tool can delegate to specialised AI provers
 such as \aristotle~\cite{Aristotle25}. The agent operates autonomously within a
 fixed token budget.

In principle, the PBT-based checks from earlier stages (specification
validation, invariant inference) could be all replaced by a loop over
AI-assisted interactive proof attempts. We experimented with this alternative
and found it impractical: even for simple tasks, the token and time costs of
attempting formal proofs vastly exceeded those of PBT, which provides the same
filtering effect at a fraction of the cost. This confirms our design choice of
reserving interactive proving for the final stage, where it is strictly
necessary.


\vspace{-5pt}

\section{\tool Testing Infrastructure}
\label{sec:pbt-filter}

This section details the PBT machinery that underpins the pipeline stages
described in \autoref{sec:overview}: generating specification tests, handling
existential quantifiers in statements, testing programs and invariants, and the
role of \lean meta-programming.
All PBT checks in \tool are built on \plausible~\cite{plausible}, a
property-based testing framework for \lean~4 that integrates with the tactic
system. Given a theorem statement, \plausible generates random inputs and
attempts to refute the theorem by finding counterexamples.

\vspace{-5pt}

\subsection{Handling Existential Quantification}
\label{sec:plausible-extensions}

PBT fundamentally struggles with existential quantification. Universally
quantified properties can be tested by \emph{sampling} inputs and checking the
predicate, but negating an existential postcondition requires showing that
\emph{no} witness exists---which in general entails reasoning over the entire
domain. This pattern is pervasive in verification conditions for array
programs. For example, to test
{\abovedisplayskip=4pt\belowdisplayskip=4pt
\[(\forall i,\, i < \mcode{arr.size} \to \mcode{arr}[i] \ge 0) \to \mcode{sum}(\mcode{arr}) \ge 0,\]}%
one must handle the logically equivalent form
{\abovedisplayskip=4pt\belowdisplayskip=4pt
\[(\exists i,\, i < \mcode{arr.size} \wedge \mcode{arr}[i] < 0) \vee \mcode{sum}(\mcode{arr}) \ge 0.\]}%
(The equivalence follows from rewriting the implication $A \to B$ as $\neg A
\vee B$ and pushing the negation through the universal quantifier.)
\plausible cannot test the existential disjunct directly, since
refuting it would require enumerating the entire domain of~$i$.

We exploit a structural property of the verification conditions that arise in
practice: existential variables typically admit finite bounds inferable from the
theorem structure. In the example above, the constraint $i < \mcode{arr.size}$
bounds the existential variable~$i$, so the disjunct can be checked by
enumerating $i \in [0, \mcode{arr.size})$.
We extend \plausible with a lightweight heuristic
that extracts subexpressions appearing in inequality constraints as candidate
bounds for existential variables and uses \lean tactics (\eg, \code{grind},
\code{omega}) to discharge the resulting bound obligations---\ie, to prove that
the candidate expression is indeed an upper bound, which is necessary for the
enumeration to be sound.
When bounds can be
established, existential quantification reduces to bounded enumeration during
testing. While this does not solve the challenge of testing arbitrary
existential quantifiers, it covers the common patterns that arise in program
verification and is effective in practice.

\vspace{-5pt}

\subsection{Testing Specifications}
\label{sec:spec-pbt}

\autoref{sec:specsyn} described the three PBT checks---pre/postcondition
soundness, and output uniqueness---that \tool generates for each test case. Here
we formalise the underlying definitions.
Let $I$ denote a set of inputs, let $\psi(i)$ be the precondition, and let
$\varphi(i, o)$ be the postcondition. For an input $i \in I$, let
$\widehat{O}(i)$ denote the set of intended outputs. A postcondition $\varphi(i,
o)$ is \emph{precise} on $I$ if it satisfies:
\begin{enumerate}[topsep=2pt,itemsep=0pt,parsep=0pt]
  \item \emph{Soundness}: all intended outputs are accepted, \ie,
    $\forall i \in I,\, \forall o \in \widehat{O}(i),\, \varphi(i, o)$.
    A violation means the spec is too strong.
  \item \emph{Completeness}: all unintended outputs are rejected, \ie, $\forall
    i \in I,\, \forall o \notin \widehat{O}(i),\, \neg\varphi(i, o)$. A
    violation means the spec is too weak.
\end{enumerate}
During spec generation, the LLM produces concrete test cases alongside the spec.
Each test case consists of an input~$i$ together with a representative intended
output $\widehat{o} \in \widehat{O}(i)$. To validate soundness, for each test
case $(i, \widehat{o})$, \plausible checks (a)~the input satisfies the
precondition and (b)~the input/output pair satisfies the postcondition. Failure
in either check indicates a flaw in either the generated specification or the
tests; the failing check and counterexample are fed back to the LLM, which
revises the spec (\cf~\autoref{sec:specsyn}).

Testing completeness is more challenging, because it requires checking that
\emph{no} unintended output satisfies the postcondition---a universal statement
over an unbounded output domain.
However, for many algorithmic problems
(\eg, typical \tname{LeetCode} tasks), the intended output is deterministic:
$\widehat{O}(i) = \{\widehat{o}\}$. Under this assumption, completeness
simplifies to verifying the \emph{uniqueness} of the intended output, yielding
the testable form $\forall o \neq \widehat{o},\, \neg \varphi(i, o)$. Checking
completeness thus becomes a search for a spurious output $o \neq \widehat{o}$
that erroneously satisfies $\varphi$.
In practice, spurious outputs often share structure with the intended one. We
therefore first sample candidate outputs randomly, then apply small mutations
to $\widehat{o}$, checking via \plausible whether any candidate inadvertently
satisfies~$\varphi$. The mutations are type-directed:
Booleans are flipped; numeric types (\code{Nat}, \code{Int}, \code{Char}) are
perturbed by a small additive delta (with \code{Int} additionally supporting
negation); pairs $\alpha \times \beta$ have one component mutated (or, when
$\alpha = \beta$, swapped); collections (\code{Array}~$\alpha$,
\code{List}~$\alpha$, \code{String}) undergo element-level mutation, random
deletion, or random insertion.

All mutations are implemented via \lean \emph{type classes}---a mechanism for
expressing constrained polymorphism, similar to \tname{Haskell}'s type classes
or \tname{Rust}'s traits~\cite{leantc}. A type class \code{Mutatable}~$\alpha$
declares a single operation \code{mutate} : $\alpha \to$ \code{Gen}~$\alpha$,
where \code{Gen} is \plausible's randomised generation monad. The operation
takes a value and returns a ``nearby'' variant, enabling corpus-guided fuzzing
on top of \plausible.
Concrete instances provide the implementation for each base type listed above.
Composite types such as multi-dimensional arrays are supported automatically
through recursive instance resolution: the instance for \code{Array}~$\alpha$
delegates element-level mutations to the \code{Mutatable}~$\alpha$ instance,
which in turn may recurse further.

\vspace{-5pt}

\subsection{Testing Programs and Invariants}
\label{sec:proginvtesting}

We adapt PBT to test both synthesised programs and their loop invariants, as
outlined in \autoref{sec:progsyn}, by transforming a \velvet method into a
testing harness that interleaves execution with checks done at runtime.
\autoref{fig:pbt-testing-procedure} illustrates the result for our running
example.

\begin{figure}[t]
\centering
\begin{minipage}[t]{0.95\linewidth}
\begin{leancodesmall}[highlightlines={3,9-11,17-19,21-22}, highlightcolor=highlightsnippet]
|\lw{method}| CheckSortedAndRotatedTesting
do
  nums |$\leftarrow$| sample(|\texttt{Array Int}|) satisfying precondition
  let n := nums.size
  if n |$\leq$| 1 then |\lw{return}| true
  else
    let |\lw{mut}| drops : Nat := 0
    let |\lw{mut}| i : Nat := 0
    if not (plausible_test "inv_bounds") then 
      |\lw{throwError}| "inv_bounds failed at entry"
    -- (other invariants checked similarly)
    |\lw{while}| i < n do
      let a := nums[i]!
      let b := nums[(i + 1) % n]!
      if b < a then drops := drops + 1
      i := i + 1
      if not (plausible_test "inv_bounds") then 
        |\lw{throwError}| "inv_bounds not preserved"
      -- other invariants are checked similarly
    let result : Bool := drops |$\leq$| 1
    if not (plausible_test "post") then 
      |\lw{throwError}| "postcondition failed"
\end{leancodesmall}
\end{minipage}
\setlength{\abovecaptionskip}{10pt}
\setlength{\belowcaptionskip}{-15pt}
\caption{A testing procedure for \code{CheckSortedAndRotated} }
\label{fig:pbt-testing-procedure}
\end{figure}

The harness has three noteworthy components (highlighted in
\autoref{fig:pbt-testing-procedure}).
First, the input is sampled randomly subject to the precondition (line~3):
\plausible generates candidate arrays and retains only those satisfying
\code{precondition}.
Second, each loop invariant is checked both at loop entry (lines~9--11) and
after every iteration (lines~17--19). A failure pinpoints the exact invariant
and the concrete input that violates it, giving the LLM targeted feedback for
revision.
Third, the postcondition is checked on the final result (lines~21--22),
catching implementations that compute an incorrect answer despite maintaining
all invariants.

\begin{table*}[t]
\centering
\footnotesize
\setlength{\abovecaptionskip}{5pt}
\setlength{\belowcaptionskip}{-15pt}
\caption{Specification issues discovered in published benchmark suites for vericoding in \lean.}
\label{tab:spec-issues}
\begin{tabular}{llcp{9cm}}
\toprule
\textbf{Benchmark} & \textbf{Issue type} & \textbf{\#} & \textbf{Instances} \\
\midrule
\multirow{2}{*}{\verina~\cite{YeYHKYS25}}
  & Underspecified postcondition & 12 & adv.8, adv.47, adv.71, bas.22, bas.34, bas.37, bas.46, bas.77, bas.84, bas.95, bas.97, bas.103 \\
  & Incorrect postcondition      & 4  & adv.10, adv.12, adv.13, bas.79 \\
\midrule
\multirow{3}{*}{\clever~\cite{ThakurLTSZZDYC25}}
  & Underspecified postcondition     & 16 & P7, P24, P27, P29, P30, P44, P49, P68, P69, P79, P84, P88, P89, P96, P156, P161 \\
  & Implementation issue   & 1  & P9 \\
  & Possible incorrect specification & 1  & P94 \\
\bottomrule
\end{tabular}
\end{table*}

\subsection{Using \lean Meta-Programming}
\label{sec:metaprogramming}

The testing harness of \autoref{fig:pbt-testing-procedure} must inspect a
method's structure to instrument each invariant, so it cannot be an ordinary
function. We implement it via \lean's meta-programming
facilities~\cite{metalean}, which manipulate program components during
compilation.
\velvet is shallowly embedded in \lean: a method is a monadic computation in
\code{VelvetM}~\cite{loom-paper}, modelling mutable state and imperative
control flow. We enrich it with two monad
transformers~\cite{LiangHJ95} for failure reporting and randomised input
generation:

{\abovedisplayskip=18pt\belowdisplayskip=8pt
\!\!\!\!\!\!\code{def VelvetTestingM := ExceptT String (StateT StdGen VelvetM)}
}

\vspace{2pt}
\noindent
Here, \code{ExceptT String} adds the ability to abort with an error message
(used when a check fails), and \code{StateT StdGen} threads a pseudo-random
generator through the computation (used by \plausible for sampling).
During \emph{elaboration}---the phase in which \lean resolves implicit
arguments, synthesises type class instances, and type-checks terms---our
metaprogramming code embeds every \code{VelvetM} operation into
\code{VelvetTestingM} (via the standard monad transformer \code{lift}) and
replaces each invariant annotation with an inlined \plausible check. 
The result is exposed as the \code{velvet_plausible_test} command
(\cf~\autoref{fig:velvet-synth}), which runs property-based tests on \velvet
methods and reports any violations and counterexamples.



\vspace{-5pt}

\section{Specification Inference Evaluation}
\label{sec:eval-specs}

Specification quality is the bedrock of the vericoding pipeline: a flawed spec
renders every subsequent stage vacuous. We evaluate how faithfully LLM-generated
specifications capture problem intent and whether PBT can serve as a quality
oracle, which is \emph{a priori} not obvious, since specs are logical formulas
whose counterexamples may be hard to find by random sampling. We test against
the \verina benchmark~\cite{YeYHKYS25}; the results are surprising---both in
accuracy and in what PBT reveals about existing benchmarks.

\vspace{-7pt}

\subsection{Assessing Specification Accuracy}
\label{sec:rq1}

\begin{rqbox}
\textbf{RQ1}: How well do the specifications generated by \tool capture the
intent of the natural language description?
\end{rqbox}

\noindent
To answer this question, we need a benchmark with ground-truth specifications
against which we can compare. We choose \verina, the largest available
\lean-based vericoding benchmark to date~\cite{YeYHKYS25}, which pairs 189
natural-language problem descriptions with manually curated formal specs. We
excluded one problem (\code{basic_104}) from our experiment because its
reference specification relies on a custom data structure that prevents
automated comparison.

For each of the remaining 188 problems, we run our specification generator
(\autoref{sec:specsyn}) to produce a candidate $(\mathit{pre}_{\text{gen}},
\mathit{post}_{\text{gen}})$ and compare it against the reference
$(\mathit{pre}_{\text{ref}}, \mathit{post}_{\text{ref}})$.
We assess semantic accuracy via formal equivalence checking in \lean:
(i)~the preconditions are equivalent,
$\mathit{pre}_{\text{gen}} \leftrightarrow \mathit{pre}_{\text{ref}}$, and
(ii)~under the precondition the postconditions agree,
$\mathit{pre}_{\text{gen}} \rightarrow (\mathit{post}_{\text{gen}}
\leftrightarrow \mathit{post}_{\text{ref}})$.
We discharge these obligations using \aristotle AI prover~\cite{Aristotle25},
which was freely available at the time of submission, though high demand for the
system led to long waiting times. 
When equivalence cannot be established, we manually inspect the specifications
to categorise the discrepancy.
The 188 problems break down as follows:
\begin{enumerate}[topsep=2pt,itemsep=1pt,parsep=0pt]
  \item \emph{Equivalent (150):} generated specifications are logically
    equivalent to the reference ones.
  \item \emph{Reference issues (16):} inconsistencies between the benchmark's
    reference postconditions and the intended semantics
    (\cf~\autoref{sec:spec-bugs}).
  \item \emph{Precondition variations (14):} preconditions differ slightly (\eg,
    admitting broader input sets) but remain valid semantic extensions. These do
    not affect correctness of synthesised programs.
  \item \emph{Ambiguous language (3):} vague descriptions admit multiple
    interpretations, leading to divergent but defensible specifications.
  \item \emph{Over-restrictive postconditions (2):} our postconditions impose
    unstated constraints to ensure determinism (\eg, a tie-breaking rule not
    mandated by the problem). While stricter than necessary, these specs are
    still sound and lead to correct, if slightly constrained, implementations.
  \item \emph{Incorrect specifications (3):} the specifications generated by
    \tool are semantically inconsistent with the problem: incorrect
    preconditions (2) and an incorrect postcondition (1).
\end{enumerate}
Categories~(1)-(4) are correct or defensible specifications (169/188, 89.9\%).
The over-restrictive cases~(5) stem from a preference for deterministic
specifications; while stricter than necessary, they are still sound. Only
category~(6) represents genuine errors: our approach deviates from the intended
semantics in just 5/188 instances (2.6\%)---lower than the inconsistency rate of
the human-written benchmark itself (16/188, or 8.5\%, category~(2),
\cf~\autoref{sec:spec-bugs}).
The PBT techniques from \autoref{sec:spec-pbt} played a critical role: PBT
caught errors in two specifications that had already passed the LLM judge,
reducing incorrect specifications from 7~(3.7\%) to 5~(2.6\%). It also uncovered
one case where the generated test cases did not satisfy the stated preconditions
despite passing the judge.

The three incorrect specs (category~6) are all \emph{over-constraining} output
in some edge cases: a spurious frequency bound, and an input bound mistakenly
imposed on the output. Because the generated test cases happen to satisfy these
constraints, PBT cannot detect them---a fundamental limitation of testing-based
validation.

\vspace{3pt}
\noindent
\textbf{Answer to RQ1.}
\tool's specification generator matches the intent of the natural-language
description in 97.4\% of cases. PBT filtering caught 2 of the 7 errors that pass
the LLM judge.

\vspace{-6pt}

\subsection{Specification Defects in Benchmarks}
\label{sec:spec-bugs}

Beyond evaluating our own specifications, we applied specification-level PBT to
the reference specs of both \verina and \clever~\cite{ThakurLTSZZDYC25}. 
For each benchmark problem, we run the three PBT checks of
\autoref{sec:spec-pbt}---precondition soundness, postcondition soundness, and
output uniqueness---using the test cases provided by the benchmarks. 


Without any LLM assistance, this process uncovered 13 issues in \verina and 18
in \clever within one day on a single Apple M4 MacBook Pro (14-core CPU, 24~GB
RAM). Specification-level PBT is dramatically more cost-effective than LLM-based
equivalence checking in \lean: the 13 \verina issues require only 9 minutes of
PBT, whereas \aristotle-based equivalence checking takes approximately 8 hours
(both are run with GPT-5.2 at a USD \$1 budget per problem, though most problems
consume less than \$0.2).

In total, we identify 16 issues in \verina (out of 189 specifications, 8.5\%)
with the help of \aristotle, and 18 in \clever (out of 161 specifications,
11.2\%) purely via PBT, summarised in \autoref{tab:spec-issues}.
Although our primary focus was testing specs, sampling inputs and executing
reference implementations can also surface implementation bugs (\eg, problem P9
in \clever).
We reported all findings to the benchmark authors. The \verina authors
acknowledged and addressed 15; the remaining one is under review. The \clever
authors acknowledged the 18~issues but have not yet released fixes.


\vspace{-5pt}

\section{Certified Program Synthesis Evaluation}
\label{sec:evaluation}

We now present our evaluation of the \tool synthesis pipeline on a new benchmark
of imperative-style problems (\autoref{sec:benchmark}), comparing the
multi-modal \tool pipeline against a single-mode \lean baseline
(\autoref{sec:rq2}), assessing the provability of residual obligations
(\autoref{sec:rq3}), and testing robustness across different LLMs
(\autoref{sec:rq4}).

\vspace{-5pt}

\subsection{Benchmark Selection}
\label{sec:benchmark}

Existing \lean benchmarks (\verina~\cite{YeYHKYS25},
\clever~\cite{ThakurLTSZZDYC25}) favour functional patterns (\eg, recursive
list traversals), omit complexity constraints, and risk training-data
contamination due to long public availability.
We therefore construct a benchmark of 50 \leetcode problems targeting
imperative algorithms (two-pointer, sliding window, random index access).
We manually selected 15 problems during development; 35 more were sampled
using Claude Opus~4.6 requiring:
simple input types, no complex data structures, deterministic specifications,
a mix of classical and recent problems, and easy-to-medium difficulty.
For each problem, Claude suggested a target time complexity based on input
constraints; we manually verified all annotations and applied minor
simplifications where needed. Below, we report results on all 50 problems,
distinguishing the development and evaluation sets where relevant (\eg,
\autoref{tab:synthesis-results}).

\vspace{-5pt}

\subsection{Multi-Modal vs. Single-Mode Synthesis}

\label{sec:rq2}

One might expect that handing a formal specification to a frontier LLM and
asking it to produce executable \lean---without staging through an imperative
DSL or inferring invariants---would suffice. We call this \emph{single-mode
synthesis}: the LLM directly produces a \lean definition with its proof, verified
solely by the type checker. Does \tool's staged pipeline offer a
measurable advantage?

\begin{rqbox}
\textbf{RQ2}: Does the \tool pipeline produce more certified solutions than a
single-mode \lean baseline at the same cost?
\end{rqbox}

\begin{figure}[t]
  \centering
  \includegraphics[width=\linewidth]{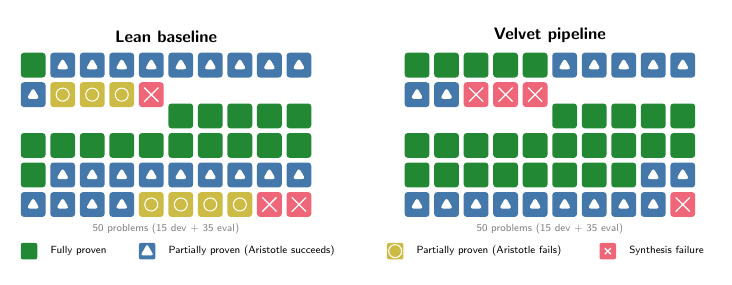}
  \setlength{\abovecaptionskip}{-5pt}
  \setlength{\belowcaptionskip}{-10pt}
  \caption{Synthesis results for single-mode \lean and \velvet.}
  \label{tab:synthesis-results}
\end{figure}

\vspace{3pt}
\emph{Setup.}~~
We compare the full \tool pipeline (\velvet code $+$ invariants $+$ multi-modal
proof) against a single-mode \lean baseline in which the LLM directly
synthesises \lean programs verified by the same prover. Both configurations use
the same LLM (GPT-5.2) and the same specifications produced in
\autoref{sec:eval-specs}; the USD \$5-per-problem budget covers only code
synthesis and proving, excluding specification generation (which is identical
for both). \autoref{tab:synthesis-results} shows the results; the development
set (15~problems, top part) and evaluation set (35~problems, bottom part) are
visually distinguished.

\vspace{3pt}
\emph{Results.}~~
For each problem we record one of three outcomes.
\emph{Fully proven:} the pipeline synthesises an implementation (with loop
invariants for \velvet) and produces a complete machine-checked correctness
proof. Overall, \velvet fully proves 28/50 problems, significantly outperforming
single-mode \lean with 17/50. (Development: 5/15 vs.\ 1/15; Evaluation: 23/35
vs.\ 16/35.)
\emph{Partially proven:} an implementation is synthesised but the proof is
incomplete---some verification conditions remain open after GPT-5.2 exhausts its
budget. In \autoref{sec:rq3} we further attempt to discharge these residual
obligations with \aristotle; here we report both sub-categories together.
\velvet yields 18 partially proven cases, while single-mode \lean produces 30.
(Development: 7 vs.\ 13; Evaluation: 11 vs.\ 17.)
\emph{Synthesis failure:} no implementation passes the pipeline's checks.
\velvet fails on 4 problems, compared to 3 for single-mode \lean. (Development:
3 vs.\ 1; Evaluation: 1 vs.\ 2.) We attribute the higher failure rate
of \velvet to the stricter validation imposed by invariant checking.

\vspace{3pt}
\emph{Overlap analysis.}~~
\autoref{tab:proof-comparison} breaks down the per-problem overlap. For example,
the entry in the first row, second column indicates that 13 problems are
fully proven by \velvet but only partially proven by \lean. While \velvet
outperforms \lean overall, it does not uniformly dominate: 4 problems are fully
proven by \lean but only partially proven by \velvet.
A representative example is \leetcode~917 (``reverse only the English letters in a
character sequence''). 
The idiomatic functional recursive implementation in \lean naturally handles
skipping non-letter characters, whereas the imperative two-pointer
implementation in \velvet requires maintaining a complex invariant,
significantly complicating the proof.
Depending on the problem, one style may yield a more concise implementation
whose correctness is easier to establish. Since \velvet is itself embedded in
\lean, users can freely fall back to single-mode synthesis.

\vspace{3pt}
\noindent
\textbf{Answer to RQ2.}
At the same USD \$5 budget, the multi-modal \tool pipeline produces 44\% more
fully certified solutions on the evaluation set (23 vs.\ 16) and 65\% more
overall (28 vs.\ 17) compared to the single-mode \lean baseline using GPT-5.2.
The advantage stems from staging: invariant inference and PBT filtering resolve
issues early, leaving the prover with fewer obligations.

\vspace{-5pt}

\begin{figure*}[t]
  \centering
  \includegraphics[width=0.42\textwidth]{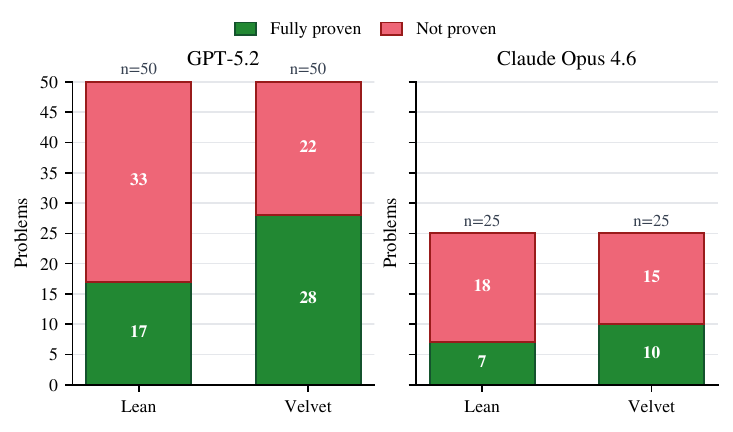}\hfill
  \includegraphics[width=0.48\textwidth]{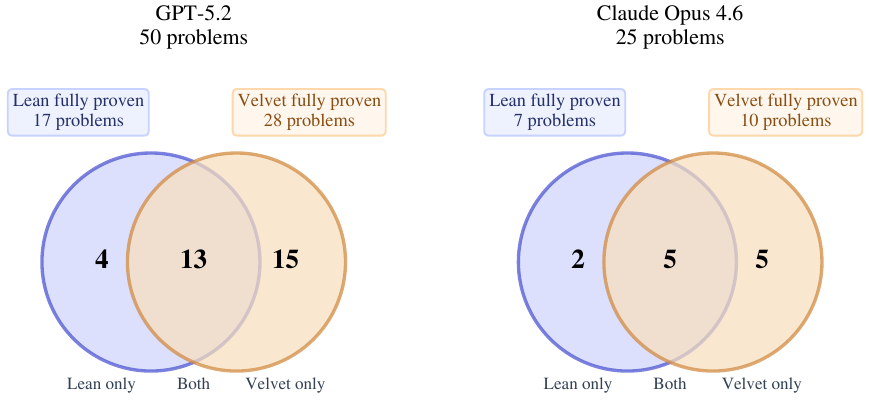}
  \caption{LLM robustness. Left: fully proven vs.\ failure for single-mode \lean
    and \velvet. Right: overlap of fully proven problems.}
  \label{fig:backend-robustness}
\end{figure*}

\subsection{Last-Mile Interactive Proof Effort}
\label{sec:rq3}

The ``partially proven'' category in \autoref{tab:synthesis-results} contains
programs whose proofs are incomplete: some verification conditions remain open
after GPT-5.2 exhausts its \$5 budget. A failed proof does not necessarily mean
the program is wrong: it may simply need more proving effort (perhaps, by a
human prover). We therefore ask whether \tool indeed produces correct programs
in these cases.

\begin{rqbox}
\textbf{RQ3}: Are partially proven  \tool-synthesised programs actually correct,
\ie, can their remaining verification conditions be discharged with additional
proof effort, and does the multi-modal decomposition make them easier to close?
\end{rqbox}

\noindent
To answer this, we attempted to close remaining verification conditions
using \aristotle~\cite{Aristotle25}, a more powerful AI prover.
For \velvet, all 18 partially proven programs are fully discharged by
\aristotle. This confirms that the \tool pipeline synthesises correct
implementations and invariants in every case; the incomplete proofs reflect
budget limitations, not errors in the synthesised artefacts.

For single-mode \lean, \aristotle successfully discharges 23 of the 30 partial
solutions but fails on the remaining~7. Notably, 1 of these problems is fully
solved by \velvet, and for the other six the corresponding \velvet verification
conditions are fully discharged by \aristotle. This suggests that
the staged pipeline generates proof obligations that are structurally easier to
close.

\vspace{3pt}
\noindent
\textbf{Answer to RQ3.}
Every partially proven \velvet program from our benchmark suite is provably
correct: the \aristotle prover successfully closes all residual obligations. 
The single-mode \lean baseline leaves seven programs whose VCs resist even
\aristotle, whereas the equivalent \velvet VCs are discharged, confirming that
multi-modal decomposition yields more tractable proof obligations.

\vspace{-6pt}

\begin{table}[t]
\centering
\setlength{\abovecaptionskip}{5pt}
\setlength{\belowcaptionskip}{-15pt}
\caption{Overlap between \velvet and single-mode \lean.}
\label{tab:proof-comparison}
\resizebox{\columnwidth}{!}{%
\begin{tabular}{lccccc}
\toprule
\makecell[l]{\velvet\,/\,\lean} & \makecell[c]{Fully\\proven} & \makecell[c]{Partial /\\\aristotle ok} & \makecell[c]{Partial /\\\aristotle fail} & \makecell[c]{Synthesis\\failure} & Total \\
\midrule
Fully proven & 13 & 13 & 1 & 1 & 28 \\
Partial (\aristotle ok) & 4 & 9 & 4 & 1 & 18 \\
Partial (\aristotle fail) & 0 & 0 & 0 & 0 & 0 \\
Synthesis failure & 0 & 1 & 2 & 1 & 4 \\
\midrule
Total & 17 & 23 & 7 & 3 & 50 \\
\bottomrule
\end{tabular}
}
\end{table}

\subsection{Robustness to LLM Choice}
\label{sec:rq4}
\begin{rqbox}
\textbf{RQ4}: Are the advantages of synthesis with a multi-modal verifier over
the single-mode one we discussed in RQ2 consistent across different frontier LLM
backends?
\end{rqbox}

\noindent
We randomly sample 25 problems from our benchmark suite and replicate the
experiment using Claude Opus~4.6 in place of GPT-5.2, keeping all other settings
identical.
\autoref{fig:backend-robustness}~(left) confirms that \velvet outperforms
single-mode \lean under both backends: 28/50 vs.\ 17/50 for GPT-5.2, and
10/25 vs.\ 7/25 for Opus~4.6. The lower absolute numbers for Opus~4.6 reflect
its higher per-token cost, which leaves less room within the fixed \$5 budget.
\autoref{fig:backend-robustness}~(right) shows the overlap of fully proven sets.
Under GPT-5.2, 13~problems are solved by both pipelines, 4~only by single-mode
\lean, and 15~only by \velvet. Under Opus~4.6, the split is 5/2/5. Changing the
backend affects which individual problems are solved but does not change the
overall ranking: \velvet remains the stronger pipeline.

\vspace{3pt}
\noindent
\textbf{Answer to RQ4.}
The results are qualitatively consistent across LLMs: the \velvet-based
multi-modal pipeline yields more fully proven solutions than single-mode \lean
under both GPT-5.2 and Claude Opus~4.6, even though the margin narrows on the
Opus subset.


\section{Threats to Validity}
\label{sec:discussion}

\paragraph{Benchmark scale and selection.}
Our benchmark comprises 50 algorithmic \leetcode problems
(\autoref{sec:benchmark}), which do not cover all software classes (\eg,
concurrent or I/O-heavy systems). Of these, 15 informed pipeline design, though
no hyperparameters were tuned on them. To mitigate selection bias, we
additionally test our specification inference against two external suites
(\verina, \clever) and observe consistent results across LLMs, suggesting
structural rather than dataset-specific advantages.

\vspace{3pt}

\emph{Fixed budget and LLM evolution.}~~
The RQ2 comparison (\autoref{sec:rq2}) uses a fixed \$5 budget per problem. At
substantially higher budgets single-mode synthesis might close the gap;
conversely, at lower budgets the gap may widen. As frontier models improve,
single-mode approaches will also get stronger. However, multi-modality reduces
cost by letting cheap validation modes (testing, SMT) filter candidates before
expensive interactive proving is invoked---a cost-reduction principle that holds
regardless of model capability.

\vspace{3pt}

\emph{Implementation and external dependencies.}~~
Although the staged decomposition of vericoding is a conceptual contribution not
tied to any particular tool, our implementation relies on \velvet, \lean, and
\aristotle. The RQ3 claim that all partial \velvet proofs are dischargeable
(\autoref{sec:rq3}) depends on \aristotle's current capabilities. Porting \tool
to another multi-modal verifier would require engineering effort, but the
pipeline architecture transfers directly.



\vspace{-5pt}

\section{Related Work}
\label{sec:related}

\paragraph{AI-assisted vericoding}
A growing body of work uses LLMs to generate verified code for specific
verifiers: auto-active approaches target
\dafny~\cite{MisuLM024,SunSPB24,LoughridgeSACSSMAMT24,BaksysZBDKH25,Laurel25,DafnyPro26},
\verus~\cite{YangLMYCGHLLLYZ25,AggarwalPW25,ChenLLGYLMYDCYLXZ25}, and
\fstar~\cite{ChakrabortyEBFF25}, while interactive-side efforts focus on
\lean~\cite{SongYA24} and \coq~\cite{Cobblestone24,PALM24,Rango24}.
Several of these systems employ staged pipelines. Misu~\etal~\cite{MisuLM024}
first generate a specification, then synthesise a verified \dafny method;
Clover~\cite{SunSPB24} generates code, docstrings, and annotations separately
before cross-validating them; \tname{ATLAS}~\cite{BaksysZBDKH25} uses an
explicit two-stage spec-then-implementation loop; and
\tname{Laurel}~\cite{Laurel25} decomposes invariant inference from proof search
in \dafny. However, all these pipelines rely on a \emph{single verification
mode} (SMT) throughout every stage, leaving no fallback when the solver times
out or cannot handle a particular obligation. 
\tool differs by building on a multi-modal verifier, matching each stage to the
most cost-effective reasoning mode: PBT for specification validation, SMT~+~PBT
for invariant inference, and interactive proofs for last-mile obligations. The
reusable insight is this \emph{mode-aware decomposition}, rather than a
prompting strategy tied to one tool's annotation language.

The most closely related agentic effort is
\tname{AutoRocq}~\cite{abs-2511-17330}, which uses an LLM agent with an
iterative feedback loop to generate tactic proofs in \coq. However,
\tname{AutoRocq} addresses only the \emph{proof generation} sub-problem:
specifications and loop invariants are produced externally (\eg, by
\tname{Frama-C} verifier~\cite{KircherCKR15}).

Mukherjee~\etal~\cite{MukherjeeLD25} use an LLM in a feedback loop with
\tname{VST}~\cite{Appel-al:BOOK14} to synthesise verified C programs, but start
from a formal specification rather than natural language and target a single
verification mode: \coq proofs in Separation
Logic~\cite{Reynolds:LICS02,OHearn-al:CSL01}.
By contrast, \tool is an end-to-end pipeline covering the full vericoding task
from natural language to certified code, using multi-modal verification to match
each sub-task to the most effective reasoning mode.

\vspace{-4pt}

\paragraph{Non-LLM certified program synthesis}
Before LLMs, several systems produced programs with machine-checkable
correctness proofs using deductive synthesis or verified compilation.
\tname{Fiat}~\cite{Delaware-al:POPL15} synthesises correct-by-construction
implementations of abstract data types in \coq via tactic-driven stepwise
refinement.
\tname{Rupicola}~\cite{PitClaudel-al:PLDI22} compiles idiomatic \coq
functions to efficient imperative code, preserving correctness through the
compilation pipeline.
The work by Watanabe~\etal~\cite{WatanabeGPPS21} certifies the output of a
separation logic-based program synthesiser~\cite{Polikarpova-Sergey:POPL19} in a
post-hoc way by translating deductive derivations into \coq proofs.
More recently, Goldstein~\etal~\cite{GoldsteinPTSLP25} used deductive synthesis
in \lean to produce certified constrained random generators for property-based
testing~\cite{ClaessenH00}.
These approaches require the specification to be written manually in the
verifier's logic; \tool complements them by using LLMs to bridge the gap from
natural language to formal specifications.

\vspace{-4pt}

\paragraph{Specification quality and validation}
Ensuring generated specifications are neither too weak nor too strong is a
recurring challenge. Existing approaches rely on manual
tagging~\cite{MisuLM024}, user studies~\cite{MaL0XB25}, LLM-as-judge with manual
inspection~\cite{VericodingBenchmark25}, design disciplines such as
non-computable specifications~\cite{ThakurLTSZZDYC25}, or multi-stage evaluators
mixing theorem proving with testing~\cite{YeYHKYS25}. Notably,
\verina~\cite{YeYHKYS25} also uses PBT (via \lean's \plausible
library~\cite{plausible}) to check spec soundness and completeness, but as a
\emph{fallback} when theorem proving is inconclusive and only for retrospective
benchmark evaluation---not as an online filter during synthesis. Moreover,
\verina compares generated specs against \emph{ground-truth} reference specs,
whereas our checks require only test cases and the generated spec itself.
\tname{ATLAS}~\cite{BaksysZBDKH25} attempts a similar uniqueness check by
generating \dafny lemmas that derive a contradiction from alternative outputs,
but reports that this often exceeds SMT capabilities.
Using randomised testing to validate formal specs has a long
history~\cite{Hughes11,Bulwahn12,ParaskevopoulouHDLP15}; however, these works
target hand-written specs in stand-alone proof assistants, not LLM-generated
specs inside a synthesis pipeline.
Lahiri~\cite{Lahiri24} proposes \emph{symbolic specification testing}, checking
LLM-generated \dafny specs against concrete input/outputs via the \dafny
verifier. Our approach builds on these ideas but (1)~uses PBT as the
\emph{primary} method and as a filtering stage in the synthesis pipeline (not a
fallback or post-hoc evaluation), (2)~replaces SMT-based checking with
randomised testing, which we found more cost-effective in \lean, and
(3)~includes a uniqueness check that PBT handles naturally but SMT often cannot.

\vspace{-4pt}

\paragraph{Multi-modal verifiers}
\K~\cite{Rosu17} derives multiple reasoning modes (execution, model checking,
deductive verification) from a single semantics but lacks interactive proofs.
\ivy~\cite{PadonMPSS16,McMillanP20} and \veil~\cite{VeilPaper} combine SMT
verification with model checking and manual proofs, but target distributed
systems. \velvet~\cite{VelvetPaper26}, which \tool builds on, unifies SMT
automation, interactive \lean proofs, and PBT---making all three modes available
to an LLM-driven pipeline. To our knowledge, no prior multi-modal verifier has
been used for implemting end-to-end LLM-assisted certified synthesis.


\section{Conclusion}
\label{sec:conclusion}

We have presented \tool, an agentic pipeline for certified program synthesis
that leverages multi-modal verification---combining testing, SMT-based
automation, and interactive proof scripting---to decompose vericoding into
stages, each matched to the most effective reasoning mode.
Our evaluation shows that this decomposition yields more fully certified
solutions than single-mode baselines at the same cost, while randomised
specification testing catches defects that existing benchmarks miss.
While a sufficiently powerful AI prover could in principle subsume every stage,
our results demonstrate that reserving interactive proving for the final
stage---and delegating earlier filtering to cheaper modes such as PBT and
SMT---reduces costs by orders of magnitude without sacrificing correctness.
Multi-modality is thus not merely a convenience but a practical necessity:
principled composition of complementary reasoning modes within a foundational
framework provides a robust foundation for future vericoding systems.

\vspace{3pt}
\noindent
\textbf{Data Availability Statement.}~~
An artefact containing the implementation of \tool, the benchmark suite of
50~\leetcode problems, and the evaluation harness for reproducing the results
from \autoref{sec:eval-specs} and \autoref{sec:evaluation} is publicly
available~\cite{leetproof-artefact}.

\bibliography{references}


\end{document}